\documentclass[a4paper,12pt]{article}

\usepackage{graphicx}

\title{`Circular type' quantum key distribution}
\author{Tsuyoshi NISHIOKA
        \thanks{Phone:+81-467-41-2181, Fax:+81-467-41-2185, 
        E-mail: nishioka@isl.melco.co.jp}
        \and Hirokazu ISHIZUKA 
        \and Toshio HASEGAWA \\ 
        \small\it MITSUBISHI ELECTRIC CORPORATION, \\
        \small\it INFORMATION TECHNOLOGY R \& D CENTER, \\ 
        \small\it 5-1-1 Ofuna, Kamakura, Kanagawa 247-8501, JAPAN
        }

\begin{document}

\maketitle

\abstract{`Circular type' interferometric system for quantum key distribution
is proposed. The system has naturally self-alignment and compensation of 
birefringence and also has enough efficiency against polarisation 
dependence. Moreover it is easily applicable to multi-party.
Key creation with 0.1 photon per pulse at a rate of 1.2KHz with a
5.4\% QBER over a 200m fiber was realized.}

\vspace{2em}

Quantum key distribution is expected as one of the most important 
technology on information security in the near future and
provides two remote parties, Alice and Bob,
with a common key in private manner. 
Many groups have reported 
its realization on optical fibers or 
others\cite{BBBSS92}-\cite{MT95}. 
Almost all reports based on 
interferometric system have a problem on alignment and compensation 
of birefringence. 
Geneva group has solved this problem using Faraday mirror 
excellently\cite{ZGGHMT97,RGGGZ98}. 
Their solution is self-alignment and 
compensation, that is, the two 
optical paths constituting its interferometric system are the same physical 
path and therefore the difference between the two paths is 
compensated automatically. 

But the solution has a flaw that it is sensitive to the polarisation
dependent loss because the first and second paths are discriminated each
other by polarisation states with the polarisation 
beam splitter\cite{RGGGZ98}.
Our proposal depicted in Fig.1 adopts circular-type optics instead of 
a Faraday mirror and a polarisation beam splitter. 

\begin{figure}[htbp]
\includegraphics
[trim = 0  75 200 580, clip, width=0.9\linewidth]
{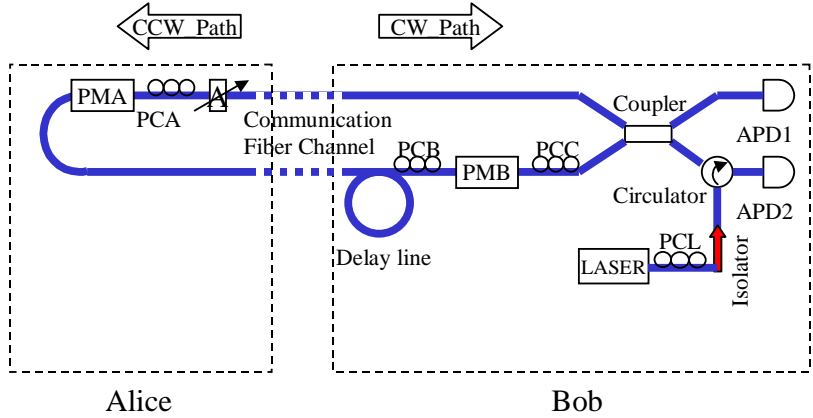}
\caption{\it Schematic Diagram of `Circular-type' quantum key 
             distribution system}
\label{fig:optics}
\end{figure}

But the optics
maintains self-alignment and compensation of birefringence because the 
light pulses circulate the same physical path each counterclockwise 
and clockwise.
The optics, then, minimizes polarisation dependent loss because the two 
paths are naturally discriminated. 
Our system 
is, in addition, simpler than the Geneva group's system 
and therefore has excellent performance on transfer efficiency.
Though the system requires two fibers as communication channel, 
the installation of two fibers is not critical because 
multi-core-fiber cables are currently used in the real world.

In Fig.1, Bob, as well as the Geneva's system, 
has a photon source and sends a photon pulse through a circulator 
into the system. 
The pulse splits at a coupler. The first half circulates the whole optics
counterclockwise, where the pulse is named CCW pulse, 
and the second half circulates clockwise, 
where the pulse is named CW pulse. 
The CCW pulse propagates
to Alice through a optical fiber (the upper fiber)
and goes back to Bob through another optical fiber (the lower fiber). 
After  passing a delay line fiber, 
the CCW pulse is applied a phase shift $\phi_{B}$
with a phase modulator in Bob (PMB) and returns the coupler.
Polarisation controllers, PCA in Alice, PCB in Bob, PCC at the 
coupler tune the pulse to a phase modulator in Alice (PMA), PMB, 
and the coupler respectively.
The CW pulse passes the PCC, PMB, PCB, and the delay line in the order and
propagates to Alice through the lower fiber. Alice does not have the
two pulses simultaneously because the delay line staggers their transit
timings.
Alice, then,
applies to the CW pulse only a phase shift $\phi_{A}$ with PMA.  
After passing PMA, the shifted CW pulse is attenuated to the average photon 
number $\mu=0.1$ per pulse with the variable attenuator A and
goes back to Bob through the upper fiber.
The returned CW pulse reaches the coupler at the same time as 
the arrival of the CCW pulse
because they travel the same optical path reversely
and the two pluses give rise to 
interference. As the results of interference, the photon is detected at 
either avalanche photo diodes APD1 or APD2 according to its phase difference 
$\Delta\phi(=\phi_{A}-\phi_{B})$. 

In practice, we have implemented the BB84 protocol\cite{BB84} using  
a SCIENTEX OPG-2000B-830 LASER with wavelength $\lambda=830nm$ and 
the repetition rate $100KHz$. The photon source is composed of the
LASER, a polarisation controller (PCL) and an isolator.
We used EG\&G SPCM-AQR-14-FCs, which are Peltier cooled 
Si-APD, as APD1 and APD2. The transmission distance between 
Alice and Bob was set 200m  and 800m fiber was used as the delay line.
We, then, obtained the high performance of 1.2Kbps as the raw creation rate 
and 5.4\% as the quantum bit error rate (QBER). 

Although almost
 all reports of quantum key distribution have targeted 
point to point link architecture, our system can, further, target one to 
many link architecture\cite{TPBB94}.
The system can naturally extend to multi-party 
quantum key distribution on looped networks
in Fig.2,
where key distribution between Bob and anyone of Alice, David, Fox, 
George, and so on is feasible. 

\begin{figure}[htbp]
\includegraphics
[trim = 80 265 150 270, clip, width=0.7\linewidth]
{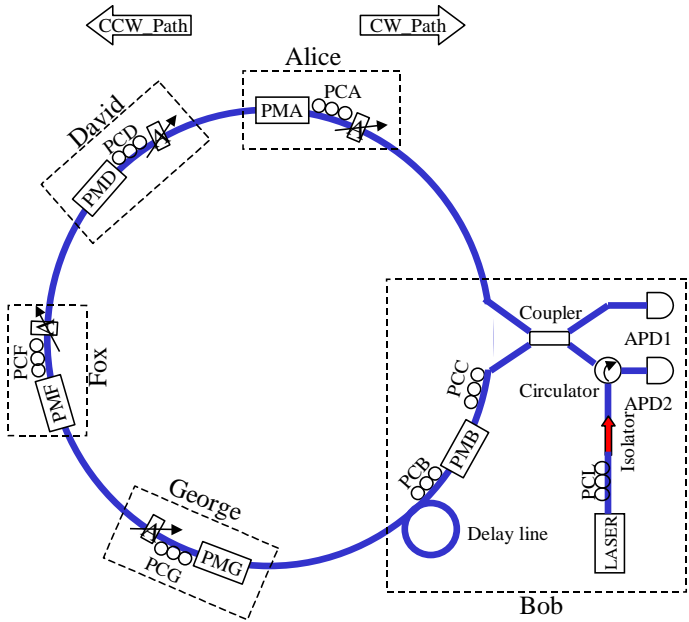}
\caption{\it Schematic Diagram of quantum key distribution on 
             looped network: Cycrography}
\label{fig:optics2}
\end{figure}

For example, Bob selects one entity from the four entities.
The selected entity and Bob, then, perform the BB84 protocol and the
other entities allow light pulses through their modules without 
disturbance.
If they disturb the pulses, Bob and his partner notice 
the disturbance as eavesdropping.
The proposed system requires only one quantum key distribution system 
though Bob can commute many entities.

In conclusion, we have presented a new simple and efficient 
interferometric system for quantum key distribution.
Our one-to-one system was tested with a raw creation
key rate 1.2KHz over a distance 200m. Our looped network system 
for more than two parties would
be tested soon.

\end{document}